# Greener processing of $SrFe_{12}O_{19}$ ceramic permanent magnets by two-step sintering

J.C. Guzmán-Mínguez [a], V. Fuertes [a,b], C. Granados-Miralles [a], J.F. Fernández [a], A. Quesada [a,*]

[a] Institute of Ceramics and Glass Materials (CSIC), Kelsen 5, 28049, Madrid, Spain
[b] Center for Optics, Photonics, and Lasers, Laval University, 2375 Terrasse street, Quebec, QC, G1V 0A6, Canada

ABSTRACT

With an annual production amounting to 800 kilotons, ferrite magnets constitute the largest family of permanent magnets in volume, a demand that will only increase as a consequence of the rare-earth crisis. With the global goal of building a climate-resilient future, strategies towards a greener manufacturing of ferrite magnets are of great interest. A new ceramic processing route for obtaining dense Sr-ferrite sintered magnets is presented here. Instead of the usual sintering process employed nowadays in ferrite magnet manufacturing that demands long dwell times, a shorter two-step sintering is designed to densify the ferrite ceramics. As a result of these processes, dense $SrFe_{12}O_{19}$ ceramic magnets with properties comparable to state-of-the-art ferrite magnets are obtained. In particular, the $SrFe_{12}O_{19}$ magnet containing 0.2% PVA and 0.6% wt $SiO_2$ reaches a coercivity of 164 kA/m along with a 93% relative density. A reduction of 31% in energy consumption is achieved in the thermal treatment with respect to conventional sintering, which could lead to energy savings for the industry of the order of $7.10^9$ kWh per year.

## 1. Introduction

Hexaferrites, particularly $SrFe_{12}O_{19}$ (SFO), cover 80% of the volume of magnets that are sold every year, reaching an annual production of 800 kilotons. This is due to the main advantages they offer that include availability, price and environmental friendliness during both mining and production [1,2]. Ferrite magnets are widely used in a variety of applications, of which electrical motors and sensors represent their largest markets [3]. As for its magnetic properties, SFO presents high coercivity and a moderate saturation magnetization [4–6]. This implies that it is often the case that remanence magnetization is the limiting factor in the energy density that a ferrite magnet can store, *i.e.* its energy product $BH_{max}$. Nevertheless, it is mandatory to achieve competitive coercivity and density values in order to maximize $BH_{max}$, and both properties strongly depend on the ceramic processing involved in the fabrication of ferrite sintered magnets [7–9].

The pre-consolidation processing of the powder is important to ensure a low degree of agglomeration of the ferrite particles for subsequent magnetic alignment and compaction. One of the key balances in hexaferrite ceramics comes from the fact that massive grain growth, which damages coercivity, is activated at the temperatures required for proper densification [5,8,10–13]. It is well know that the use of additives is crucial in tuning an appropriate microstructure. In particular, CaO and $SiO_2$ are used together to control grain growth and its impact on coercivity [11,14,15]. Recently, we have shown that it is possible to obtain dense ferrite ceramics with high coercivity using only $SiO_2$ as an additive [16,17]. Moreover, carefully designing the sintering schedule plays of course an important role as well in the final microstructure [18–20]. The most employed process at industrial level requires long dwell times, typically between 2 and 4 h, at 1200 °C and above (Fig. 1a), which has yielded traditionally high densities and moderate grain sizes in the presence of the aforementioned additives.

Given the remarkable volume of ferrite magnets that is produced every year, efforts towards reducing the dwell time associated with their sintering could lead to significant energy savings. An alternative sintering treatment, that allows reducing the dwell time during sintering treatment, is the well-known two-step sintering (TSS) process [17], with which high density values can be achieved without excessive grain growth.

In this work, we analyze the possibility to obtain high density and high coercivity $SrFe_{12}O_{19}$ sintered magnets by combining for the first time a pre-processing of the hexaferrite starting powder with a TSS

* Corresponding author.
*E-mail address:* jesus.guzman@icv.csic.es (A. Quesada).

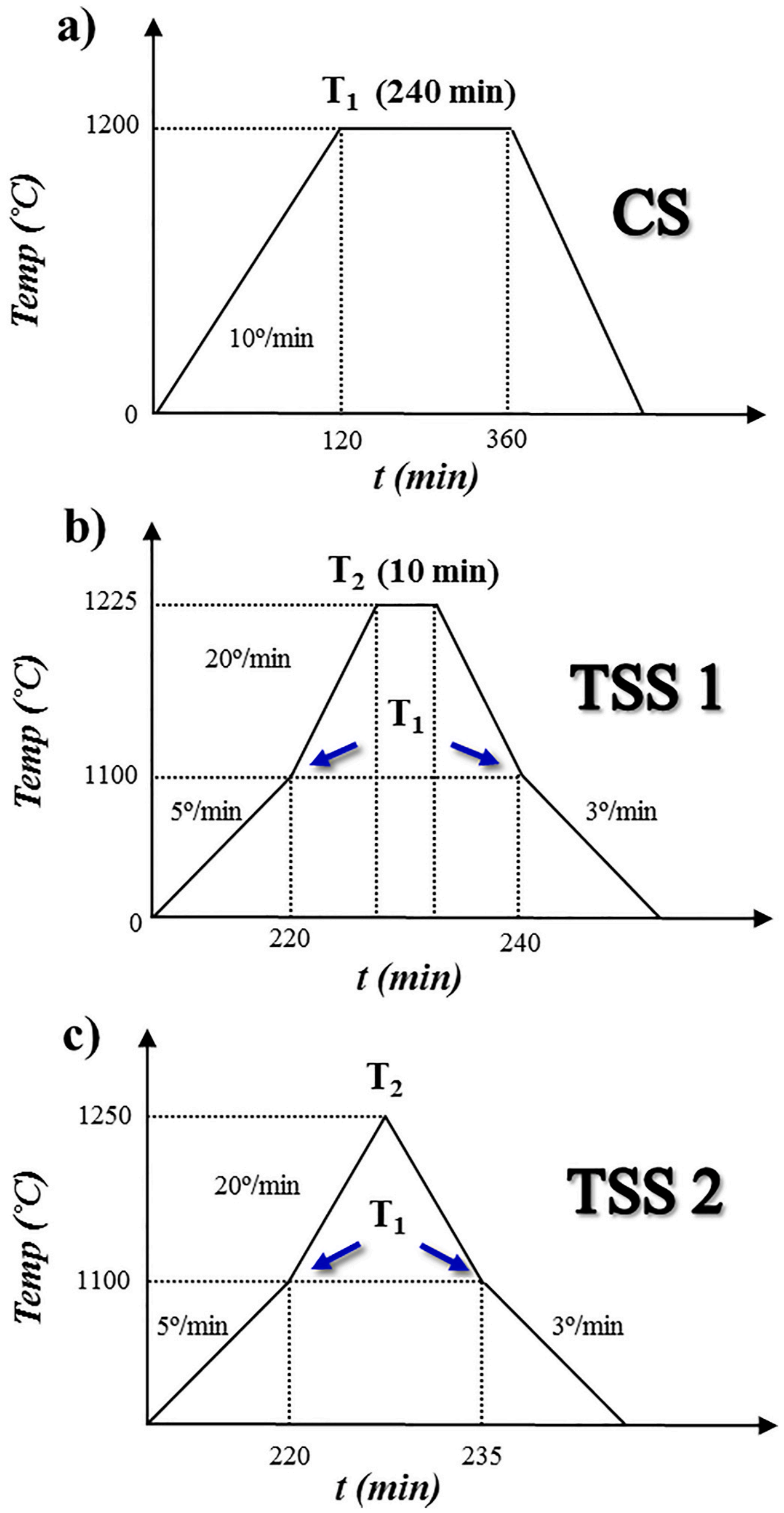

**Fig. 1.** a) Heating treatment for conventional sintering (CS); Heating treatments for two-step sintering b) (TSS 1) and c) (TSS 2).

schedule that significantly reduces the overall duration of the consolidation step.

## 2. Experimental

Firstly, 0.2 wt% of PVA (Aldrich Chemical Company, Inc.) was added to an aqueous dissolution of commercial $SrFe_{12}O_{19}$ (SFO) powders (Max Baermann GmbH 99.9%). This dissolution was homogenized and dispersed with the help of a high-shear mixer (Dissolver Simplex SL-1) by using a 5 cm in diameter cowless rotor and high solid content in water suspension, typically 60% wt, by applying 1500 rpm during 10 min.

Afterwards, different percentages of $SiO_2$ (0.2-0.4-0.6-1 %wt) were added to the SFO/PVA solution and dispersed again with the help of Disolver Simplex SL-1. The resultant mix was dried, pressed under 200 MPa and sintered by way of a two-step sintering (TSS) process. In our sintering treatment (Fig. 1b 1c), a pressed sample is heated to an intermediate temperature (T1 = 1100 °C) at 5°/min, then quickly annealed at 20°/min to a higher temperature (T2 = 1225°C-1275 °C), after a dwell between 0 and 10 min, the temperature rapidly drops (20°/min) to T1 and finally decreases to room-temperature at 3°/min. The use of this type of two-step cycle significantly reduces the sintering times from the 360 min required for conventional treatment to 240 min for TSS1 or 235 min for TSS2.

Next, we subjected selected samples ($SrFe_{12}O_{19}$ - 0.2% wt PVA with 0.6% wt of $SiO_2$) to a Spray drying process (SDP). The Spray drying method consists of two phases: 1) Particle formation: the atomization and the solvent evaporation take place and 2) Particle collection: dry and rounded particles are obtained. The rounded and dried powder obtained was pressed (while applying an external magnetic field in the final samples) and sintered by TSS.

The morphological and microstructural characterizations were carried out by means of secondary electron images of field emission scanning electron microscopy, FE-SEM (Hitachi S-4700). Powder X-ray diffraction (PXRD) data were collected in a Bruker D8 Advance diffractometer equipped with a Lynx Eye detector and a Cu target ($\lambda Cu\alpha 1$ = 1.54060 Å), after milling the sintered dense ceramics to powder. The PXRD data were fitted to a Rietveld model including $SrFe_{12}O_{19}$ as the only phase and using the software FullProf. Crystallographic densities were calculated based on the refined unit cell parameters, a and c, and assuming all atomic sites of $SrFe_{12}O_{19}$ are fully occupied by the due atoms. The relative density values were measured using by the Archimedes method and 5.1 g/cm$^3$ as the density value of $SrFe_{12}O_{19}$, which was calculated from the PXRD patterns [1]. The magnetic properties were evaluated using a homemade vibrating sample magnetometer (VSM) [21]. The magnetization curves were measured at room temperature by applying a maximum magnetic field of 1.3T.

## 3. Results

Fig. 2 shows SEM micrographs of the SFO powder before and after dispersion in high shear with PVA. As can be seen in Fig. 2a and b, the initial SFO powder contains agglomerates showing a bimodal size distribution of particles consisting of irregular small particles 200–600 nm in size and large platelet type particles 1–5 μm in size.

After mixing with PVA (Fig. 2c and d), we observe changes in the morphology and distribution of the particles. The average size of the larger particles is reduced to 1–2 μm and a more spherical shape is attained (without corners). These changes are likely the consequence of the high shear forces occurring in the Dissolver mixer. In Fig. 2d, we observe that the new agglomerates, although composed of smaller particles, appear to be arranged in larger structures. This is attributed to the encapsulation effect of adding PVA [22–24]. Higher amounts of PVA (not shown) resulted in larger agglomerates with more limited compaction capability.

Once the PVA is optimally dispersed around the SFO particles, different $SiO_2$ contents (0.2, 0.4, 0.6 and 1 %wt) were added to the mixture. As discussed in previous studies [16], the addition of $SiO_2$ has an inhibitory role on grain growth during the consolidation treatment.

SFO pellets of samples with different $SiO_2$ contents (0.2-0.4-0.6-1 %wt) and containing 0.2%wt PVA were pressed. The green bodies, with relative densities between 63 and 67% (density of SFO is 5.1 g/cm$^3$), were then subjected to various TSS processes.

The magnetization curves are shown in Fig.3. Furthermore, the values of coercivity, magnetization and relative density for each sample are collected in Table 1. As expected [11,16,25], increasing the $SiO_2$ content (%wt) tends to increase the value of coercivity ($H_c$) while magnetization at 1.3 T ($M_{1.3T}$) tends to slightly decrease. This is due to the grain growth inhibition induced by silica that lowers the average grain size, which increases $H_c$ [11]. Unfortunately, the grain growth inhibition process promotes a certain formation of secondary non-magnetic phases, such as hematite, which hinders magnetization. For the 0.2%wt $SiO_2$ content, it can be seen that, for all TSS treatments, the $H_c$ values are in the 160–168 kA/m range. For 0.4%wt and 0.6%wt, $H_c$ values are between 152 and 184 kA/m for all TSS studied. For samples with 1%wt, the $H_c$ reaches values between 152 and 192 kA/m, although at the expense of an average 6% decrease in $M_{1.3T}$ compared to the other contents, which makes us discard the 1% wt $SiO_2$ samples.

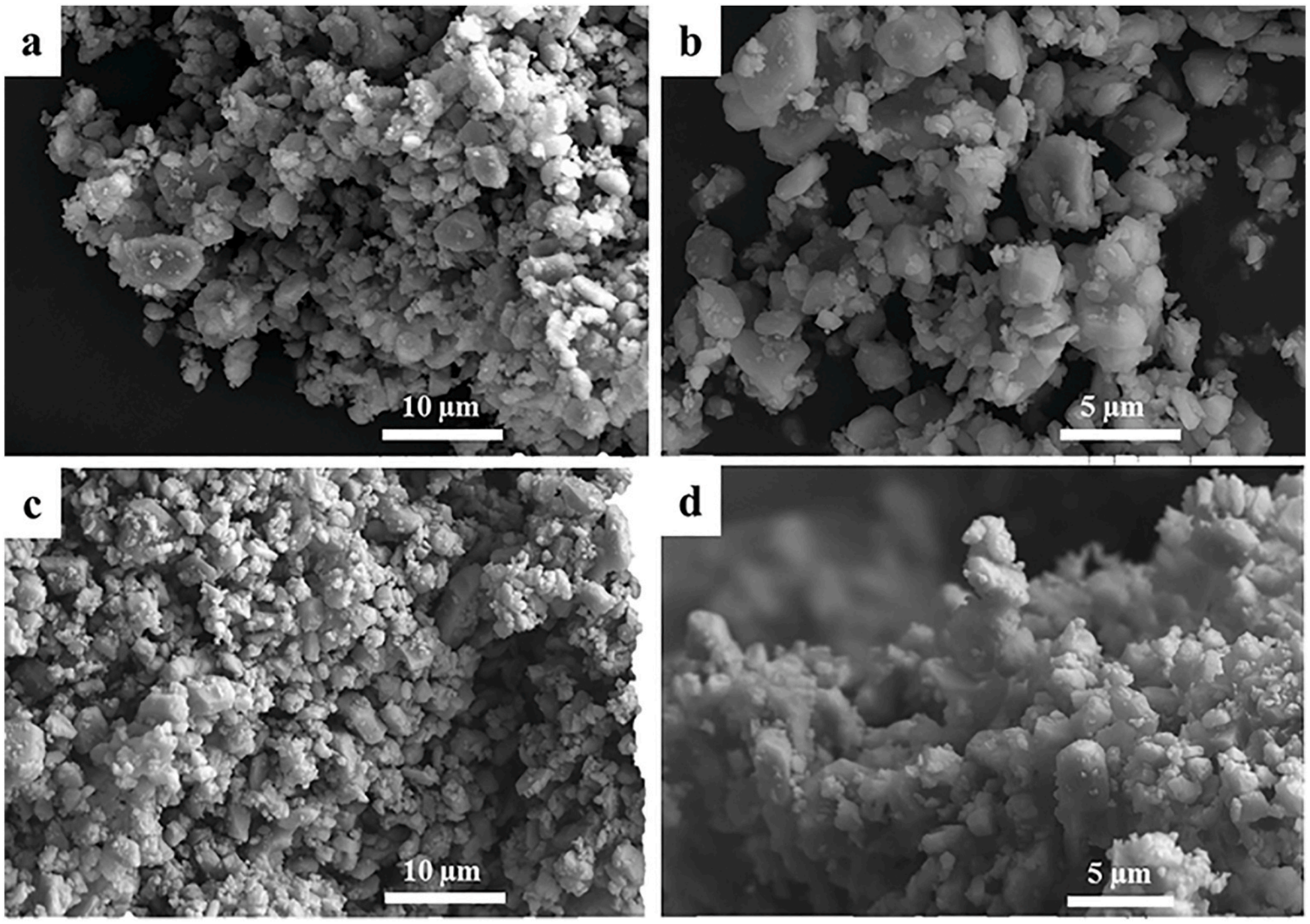

**Fig. 2.** SEM images of SFO particles (a, b) after the dispersion process without PVA and (c, d) after dispersion process with PVA.

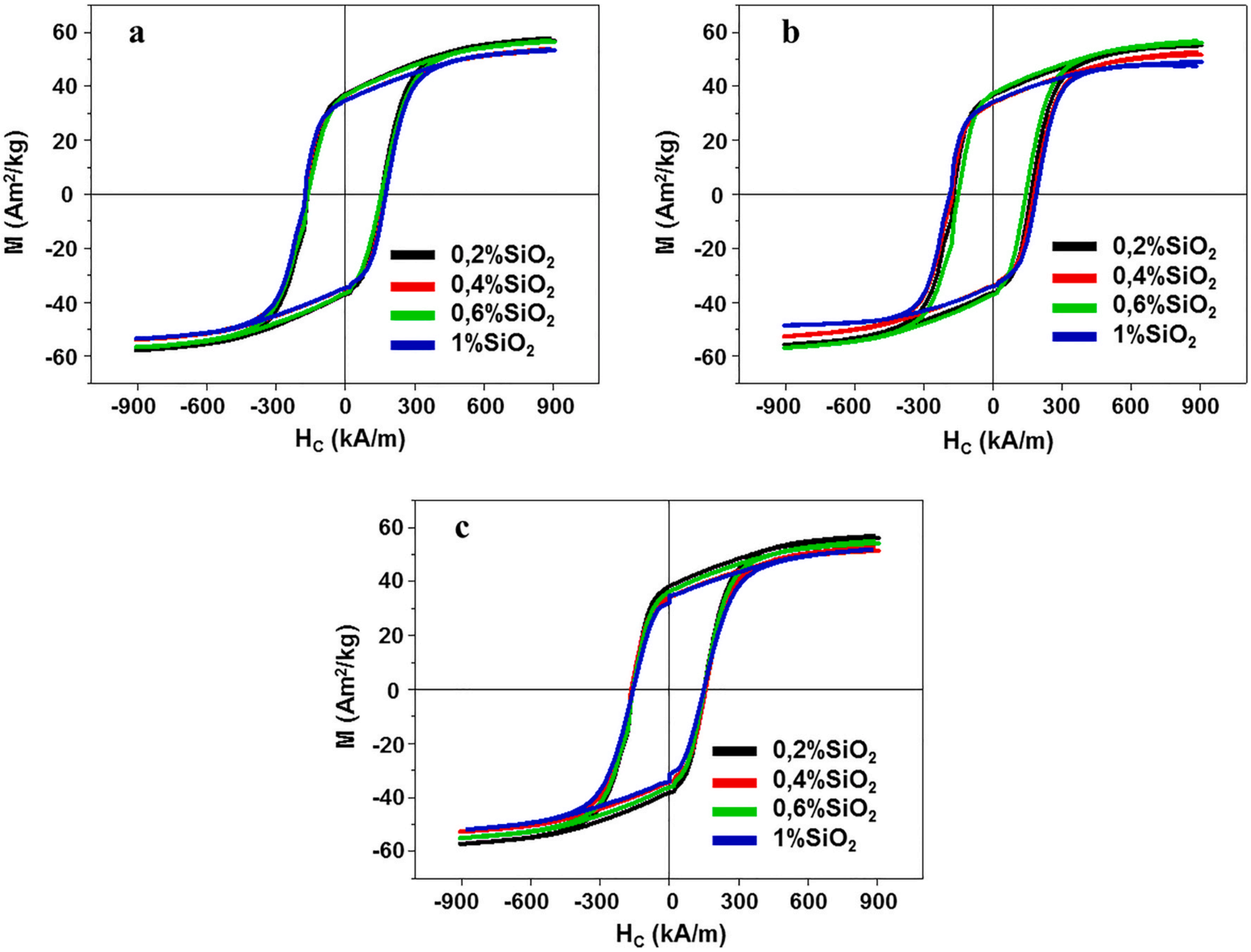

**Fig. 3.** Magnetization curves of samples with different $SiO_2$ contents sintered using different TSS-modified) processes a) 1100–1200(10 min)-1100, b) 1100–1225 (10 min)-1100 and c) 1100–1250(10 min)-1100.

**Table 1**
The table displays the values of coercivity ($H_c$), magnetization at 1.3T ($M_{1.3T}$), and relative density for all $SiO_2$ content and each TSS treatment.

| Sintering Treatment | x (% wt $SiO_2$) | $H_c$ (kA/m) | $M_{1.3T}$ ($Am^2/kg$) | Relative density (%) |
|---|---|---|---|---|
| a) 1100–1200(10 min)-1100 | 0,2 | 160 | 57 | 84 |
| | 0,4 | 168 | 54 | 90 |
| | 0,6 | 160 | 56 | 91 |
| | 1 | 176 | 53 | 90 |
| b) 1100–1225(10 min)-1100 | 0,2 | 168 | 55 | 89 |
| | 0,4 | 184 | 52 | 83 |
| | 0,6 | 152 | 56 | 92 |
| | 1 | 192 | 49 | 87 |
| c) 1100–1250(10 min)-1100 | 0,2 | 160 | 56 | 90 |
| | 0,4 | 168 | 52 | 90 |
| | 0,6 | 160 | 55 | 92 |
| | 1 | 152 | 52 | 95 |

Increasing T2 in TSS treatments increases relative density at the expense of $H$c. This competition between density and coercivity is common in SFO ceramics, as usually higher density is associated with slightly larger grain sizes. Indeed, for T2 = 1225 °C, the highest $H$c and lower densities are measured. The grain growth inhibition promoted by $SiO_2$ is a process that involves two different chemical reactions: the dissolution of Si inside the SFO grains (which is the one that actually opposes grain growth) and the decomposition of SFO onto α-$Fe_2O_3$. These two competing reactions have different kinetics and thermodynamic windows [11]. We speculate that the sintering treatment that employs T2 = 1225 °C favors the dissolution reaction over the decomposition reaction more than the other sintering schedules, leading to smaller grain sizes and thus lower density and higher coercivity. This is however not entirely true for all samples under study, which we suspect is related with the experimental error in the determination of both relative density and coercivity that we estimate in the 5–10% range. For samples treated at T2 = 1200 °C, acceptable values of $H$c between 160 and 176 kA/m are measured, although relative densities are below 91%. As we increase T2 (T2 = 1225-1250 °C) the density improves and the $M_{1.3T}$ values hardly varies. In this scenario, samples with a content of 0.6%wt of $SiO_2$, treated at T2 = 1225-1250 °C reach a reasonable compromise between the relevant magnetic and density values.

The $H_c$ values measured here are significantly higher, on average 15%, than the ones reported in our previous work using a conventional sintering procedure with a 4 h dwell time. Fig. 4 compares the microstructure of a pellet sintered at 1250 °C for 4 h and a pellet sintered using TSS with T2 = 1250 °C. For both thermal treatments, a bimodal grain size distribution can be observed, although the relative fraction of the larger grains is decreased in number and in size for TSS sintering. Moreover, the most relevant differences arise in the decrease in size of the finer grains for the TSS treatment. The average grain size determined by automatic image analysis is 3.1 ± 0.9 μm for conventional sintering and 1.1 ± 0.9 μm for TSS. Thus, a clear reduction in grain size is obtained using TSS, even for a lower $SiO_2$ content. Thus, the microstructure refinement for the TSS treatment is consequence of the grain growth control and is likely the cause for the improved $H_c$.

The coercivity of these systems is commonly described by the equation [11]:

$$H_c = \alpha H_A - N_{eff} M_s$$

Where $H_A$ is the anisotropy field of SFO, which is approximately 1.8 T [1] and $N_{eff}$ the effective demagnetization constant. However, this model, which can describe successfully systems where both coherent rotation and domain wall propagation mechanisms are occurring, is not entirely accurate for systems where magnetization reversal is dominated by domain wall propagation [4,5]. Given that the single-to-multi-domain grain size threshold for SFO is slightly below 1 μm and the grain sizes observed in Fig. 4, it seems reasonable to assume that the SFO grains of the ceramics studied here are in a multi-domain state and that domain wall propagation is dominating the demagnetization process. Under these circumstances, the main mechanism responsible for the changes in coercivity is that reducing grain size towards the single-domain threshold significantly increases $H_c$ [8].

At this stage, after choosing the appropriate additive contents ($SiO_2$ and PVA) and optimizing the TSS parameters, the effect of processing the powders with a Spray Drying Process (SDP) was investigated.

Fig. 5 shows SEM micrograph of the powders before (5a, 5b and 5c) and after (5d, 5e and 5f) the SDP. As can be seen, in Fig. 5a, b and 5c, the SFO particles present a bimodal distribution, where larger particles with sizes above 1 μm coexist with smaller ones. In addition, it is observed that the SFO particles present their characteristic platelet morphology. However, in the micrographs of the powder after SDP (Fig. 5d, e and 5f), the particle distribution is narrowed and less bimodal than for the unmodified SFO powder. The particle morphologies are slightly rounder, which may improve the final density of the samples by permitting an increased degree of packing.

On the other hand, the improvement in the dispersion of the additives ($SiO_2$ and PVA) in the ceramic matrix should be highlighted. If we focus on Fig. 5c, we see how the $SiO_2$ nanoparticles appear agglomerated on top of the SFO particles, while in Fig. 5f this does not occur. We do not observe small agglomerates in the micrographs, suggesting a successful dispersion that rends the nanoparticles not detectable at this magnification. We suggest this effect is due to the fast drying of the solutions in the spray drying chamber (typically in the range of few minutes). In the case of the powder without SDP, the drying in the laboratory stove was quite slow (in the range of several hours), causing part of the $SiO_2$ and PVA to float so that when the solution dries, it accumulates or agglomerates on the surface of the SFO particles. It should be noted here that the role that PVA plays in this process is to keep the $SiO_2$ nanoparticles dispersed on the surface of the ferrite particles. In this sense, the high shear dispersion process favors the dispersion of micro and nanoparticles. The subsequent fast drying process in the SDP ensures that the encapsulation of the ferrite particles is maintained.

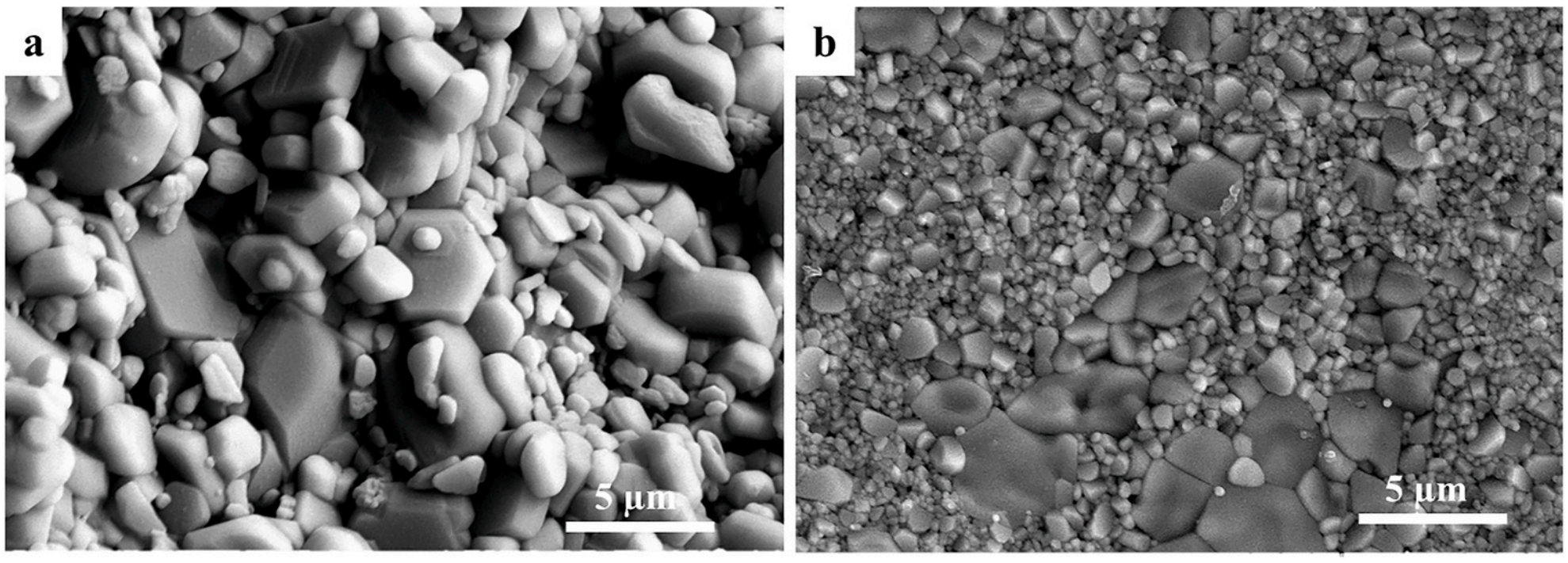

**Fig. 4.** a) SEM micrograph of a 1%$SiO_2$ SFO pellet sintered for 4 h at 1250 °C. b) SEM micrograph of a 0.6%$SiO_2$ SFO pellet sintered by TSS using T2 = 1250 °C.

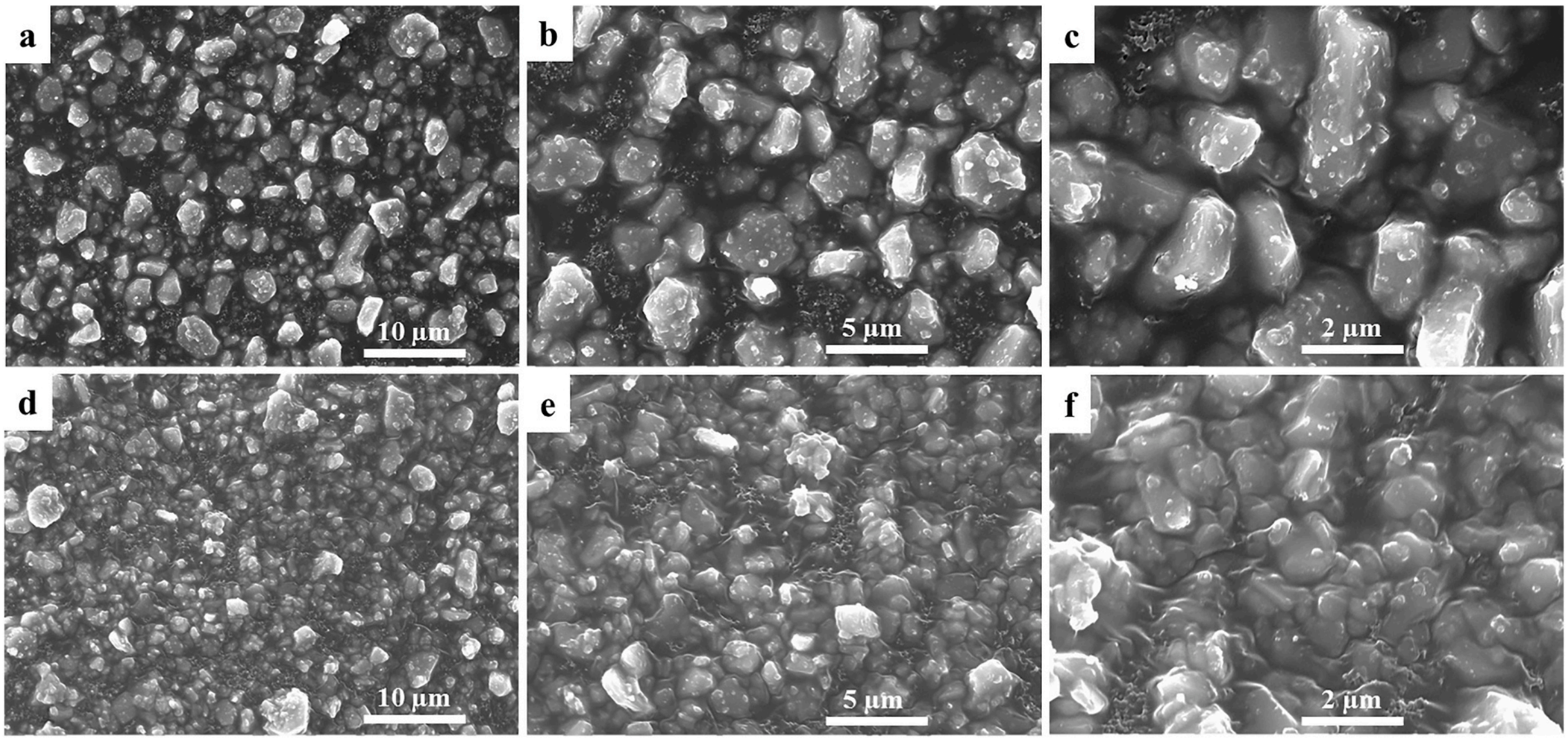

**Fig. 5.** SEM images of powder of $SrFe_{12}O_{19}$ - 0.2% PVA with 0.6% wt of $SiO_2$ before (a, b and c) and after (d, e and f) the SDP.

To further fine tune the properties, samples were then sintered using two different firing profiles: 1100–1225 [10 min dwell]-1100 will be called TSS 1 and 1100-1250-1100 will be named TSS 2. Fig. 6 shows the PXRD diffraction patterns of the SDP samples TSS1 and TSS2. Only diffraction maxima corresponding to SFO were observed, confirming the absence of secondary up to the resolution limit of the technique. In addition, the fitting of the patterns yielded a density of 5.1 $g/cm^3$, which is in agreement with the theoretical value for SFO [1].

As can be observed in Table 2, SDP samples show up to a 5% improvement in density value, reaching $\rho = 97\%$, with respect to their non-SDP counterpart.

With the aim of comparing sintered magnets fabricated following our procedure with current commercial ferrite magnets, the magnetization curves of different oriented dense pellets are represented in Fig.7, alongside that of a commercial magnet (grade Y-35).

Table 3 shows magnetic properties and relative density values of the samples and the commercial magnet. On one hand, although non-SPD samples have high coercivity values (160 kA/m), their densities are relatively low (89–90%). On the other hand, both coercivity and relative density are significantly enhanced when SPD processing the powders. In particular, the sample SFO/PVA+0.6%$SiO_2$ using SPD and the TSS 2 treatment, shows an $H_c$ value of 164 kA/m, $M_{1.3T}$ = 58 $Am^2/kg$ and a relative density of 93%. These are almost identical values as the reference state-of-the-art commercial ferrite magnet Y-35.

**Table 2**
Values of relative density (%) of samples with and without SPD, sintered using two different TSS treatments.

| Sample | Relative density (%) |
|---|---|
| SFO-PVA-$SiO_2$ SDP TSS 1 | 94 |
| SFO-PVA-$SiO_2$ Non-SDP TSS 1 | 92 |
| SFO-PVA-$SiO_2$ SDP TSS2 | 97 |
| SFO-PVA-$SiO_2$ Non-SDP TSS 2 | 92 |

Our $Mr/M_{1.3T}$ ratio is in the 0.78–0.82 range, while the commercial magnet reaches 0.93. This is likely the consequence of the fact that our orientation method, which is a delicate process, is not as optimized as that of industrial magnet manufacturers. Tokar's study demonstrated that the degree of orientation increased during sintering [26]. In particular, optimizing the viscosity of the slurred prepared with the SFO powders prior to compaction and increasing the magnetic field during the compaction process would be needed.

It is important to remark that, through the processing developed here, state-of-the-art nominal magnetic performance values are reached using a sintering schedule without dwell time. This implies a significant reduction in the energy consumption required to sinter ferrite magnets. 4 h at 1200 °C demand approximately 29 kWh, while TSS employs 20 kWh, leading to energy savings of the order of 30%. Given the 800 kilotons of ferrite magnets produced every year; industrial implementation of the sintering process presented in this work would lead to a global annual reduction in energy consumption of the order of $7 \cdot 10^9$ kWh.

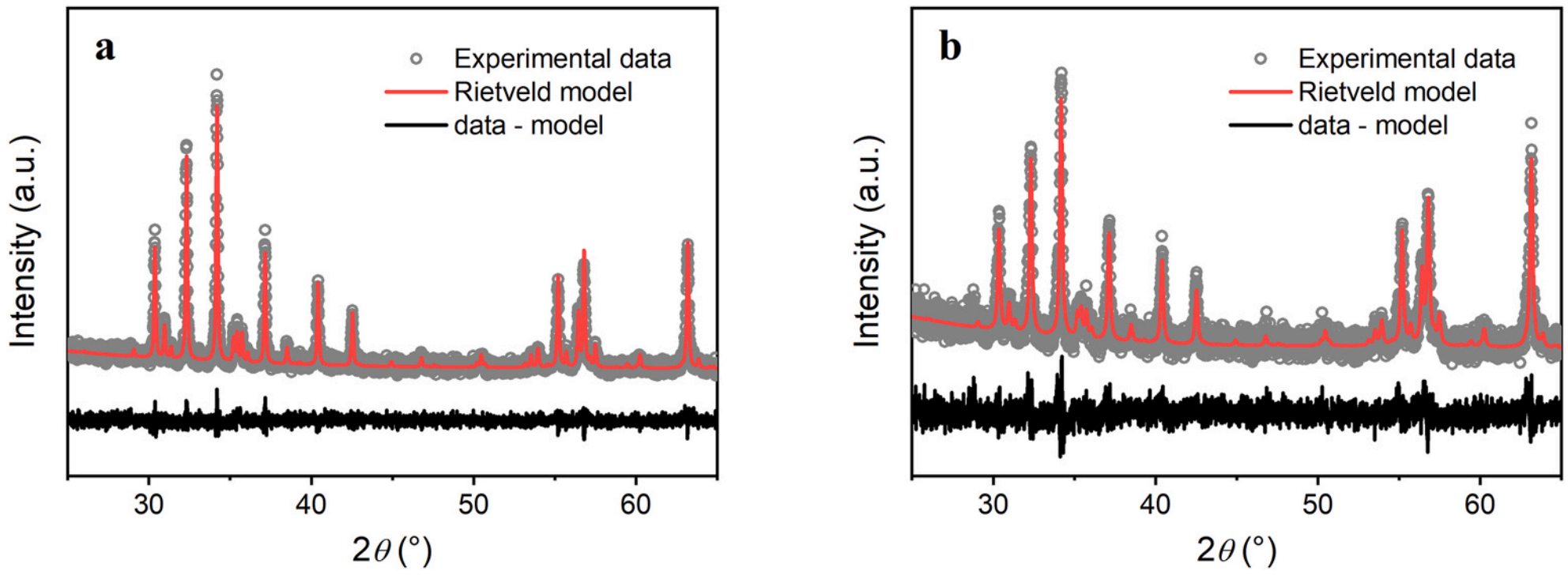

**Fig. 6.** PXRD patterns of the SSDP samples sintered using a) TSS1 and b) TSS2 schedules.

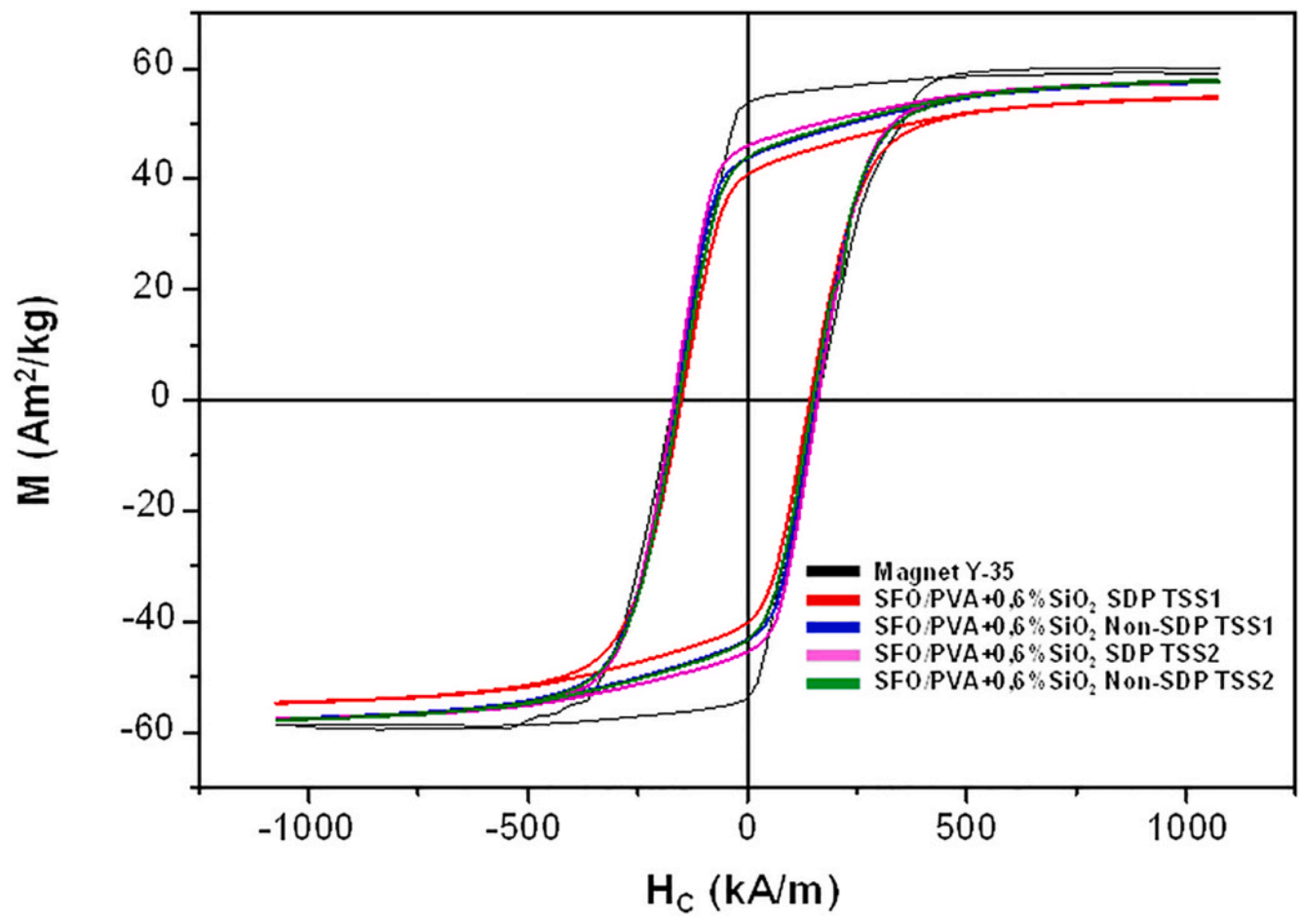

**Fig. 7.** Magnetization curves of samples of $SrFe_{12}O_{19}$ - 0.2% PVA with 0.6% wt of $SiO_2$, SPD and non-SPD sintered using a) TSS 1 and b) TSS 2.

**Table 3**
Magnetic properties and values of relative density of the oriented samples treated to two different TSS-modified treatments and of the commercial magnet Y-35.

| Sample | $H_c$ (kA/m) | $M_r$ ($Am^2$/kg) | $M_{1.3T}$ ($Am^2$/kg) | $M_r/M_{1.3T}$ (%) | Relative density (%) |
|---|---|---|---|---|---|
| Y-35 | 166 | 55 | 59 | 93 | 99 |
| SFO/PVA+0.6% $SiO_2$ SDP TSS 1 | 147 | 41 | 54 | 76 | 92 |
| SFO/PVA+0.6% $SiO_2$ Non-SDP TSS 1 | 160 | 44 | 58 | 76 | 90 |
| SFO/PVA+0.6% $SiO_2$ SDP TSS 2 | 164 | 46 | 58 | 80 | 93 |
| SFO/PVA+0.6% $SiO_2$ Non-SDP TSS 2 | 154 | 44 | 57 | 77 | 89 |

## 4. Conclusions

With the purpose of developing a greener SFO ferrite magnet manufacturing process, the sintering parameters of a TSS schedule are optimized. The influence of additives such as $SiO_2$ and PVA on the microstructure has been studied, as well as the application of the Spray drying process to modify the morphology of the SFO particles. By combining the encapsulating capacity of PVA with a high power dispersion (SDP) process, more dispersed and agglomerate-free/smaller particles are obtained. As expected, increasing the $SiO_2$ content in $SrFe_{12}O_{19}$ - 0.2%wt PVA mixtures increases the coercivity ($H_c$) value. Similarly, an increase in T2 in TSS treatments generates an increase in the relative density values. It is found that annealing up to 1100 °C at 3°/min and subsequently rapidly increasing the temperature to 1225–1250 °C before cooling back down leads to the best density and magnetic properties. The refinement of the grain size in the TSS treatment causes the notable increase of coercivity in sintered samples. Finally, the samples were oriented and compared with a commercial ferrite magnet (Y-35). Using the novel methodology proposed here for producing sintered ferrite magnets, that leads to energy savings of the order of 31%, we obtain practically the same coercivity (164 kA/m), magnetization (58 $Am^2$/kg) and relative density (93%) as for commercial ferrite magnets. Our results open the door to implementing significantly greener sintering cycles for ferrite magnets with no loss of magnetic performance.

## Declaration of competing interest

The authors declare that they have no known competing financial interests or personal relationships that could have appeared to influence the work reported in this paper.

## Acknowledgments

This work is supported by the Spanish Ministerio de Economía y Competitividad through Projects no. RTI2018-095303-A-C52 MAT2017-86450-C4-1-R, MAT2015-64110-C2-1-P, MAT2015-64110-C2-2-P, FIS2017-82415-R and through the Ramón y Cajal Contract RYC-2017-23320; and by the European Comission through Project H2020 no. 720853 (AMPHIBIAN). V.Fuertes holds a Sentinel North Excellence Postdoctoral Fellowship and acknowledges the economic support from the Sentinel North program of Université Laval, made possible, in part, thanks to funding from the Canada First Research Excellence Fund.